\documentclass[
aip,
apl,
amsmath,
amssymb,
reprint,
]{revtex4-1}

\usepackage{graphicx}
\usepackage{dcolumn}
\usepackage{bm}

\draft 

\begin{document}

\title{Free-running InGaAs single photon detector with 1~dark count per second at 10\% efficiency}

\author{B. Korzh}
 \email{Boris.Korzh@unige.ch}
\author{N. Walenta}
\author{T. Lunghi}
\author{N. Gisin}
\author{H. Zbinden}
\affiliation{GAP-Optics, University of Geneva,  Chemin de Pinchat 22, CH-1211 Geneva 4,  Switzerland }
\date{25 Feburary 2014}

\begin{abstract}
We present a free-running single photon detector for telecom wavelengths based on a negative feedback avalanche photodiode (NFAD). A dark count rate as low as 1~cps was obtained at a detection efficiency of 10\%, with an afterpulse probability of 2.2\% for 20~$\mu s$ of deadtime. This was achieved by using an active hold-off circuit and cooling the NFAD with a free-piston stirling cooler down to temperatures of -110${^o}$C. We integrated two detectors into a practical, 625~MHz clocked quantum key distribution system. Stable, real-time key distribution in presence of 30~dB channel loss was possible, yielding a secret key rate of 350~bps.
\end{abstract}


\maketitle 


Single photon detection at telecom wavelengths has attracted vast research efforts thanks to its numerous applications in quantum optics and in particular quantum key distribution (QKD) \cite{Gisin2002b, Scarani09}. An important characteristic of many tasks is that they are asynchronous, meaning the expected time of arrival of the photons is unknown, thus the single photon detectors are required to operate in the free-running regime. Superconducting nanowire single photon detectors \cite{Goltsman01} (SNSPD) generally provide the ultimate performance in terms of free-running operation with high detection efficiency \cite{marsili2013}, low dark count rates (DCR) and the absence of afterpulsing effects. Unfortunately, the need for cryogenic temperatures ($\textless$3~K) prevents their use in most applications. Another approach aiming to achieve low DCR uses frequency up-conversion to enable detection with silicon single-photon avalanche detectors (SPAD). This was recently demonstrated \cite{Shentu:13} to exhibit 25-100~cps DCR for detection efficiencies of around 10-25\%. However, this method requires a relatively complex optical setup and suffers from a narrow spectral sensitivity.

The most frequently used detectors are the InGaAs/InP SPADs because of their convenient and robust operation, compact size and low price. One drawback of these devices is the phenomenon of afterpulsing, where a spontaneous dark detection can occur shortly after a previous photon detection. This arises due to charge carrier trapping at defect sites in the SPAD multiplication region, the subsequent release of which can lead to another avalanche. Due to this, InGaAs SPADs are normally operated in the gated regime. Using short gate widths, typically less than a nanosecond \cite{walentaSine, restelli2013}, guarantees that the avalanche current is quenched very quickly, reducing the probability of a trap being filled in the first place. In order to obtain a free-running SPAD, either active or passive quenching is required \cite{cova1996}. The most effective technique for passive quenching has been demonstrated with the development of negative feedback avalanche photodiodes (NFAD) \cite{itzler2009a}. These devices have a monolithic, thin-film feedback resistor, integrated directly on the surface of the detector. Such integration reduces the parasitic capacitance, resulting in very fast passive quenching of the avalanche current, which makes straightforward operation under a DC bias possible. 

It is well understood that, beyond afterpulsing, dark-carrier generation in SPADs can occur due to either a field dependent tunnelling process or a thermally driven process \cite{Itzler2011}. Typically, SPADs have a relatively large trap-assisted tunnelling contribution, hence, it is often sufficient to only cool the devices to rather moderate temperatures, to reduce the thermal contribution well below the former effect. However, recent improvements to InGaAs material quality and careful consideration of the SPAD device structure \cite{tosi2009a}, have effectively reduced the contribution of the dark count generation through the tunnelling process. This has also been applied to NFAD design \cite{itzler2009a} and subsequent work has focused on better understanding and improving the avalanche quenching in these devices \cite{jiang2011}. Indeed, these advancements have opened up the possibility of further reducing the thermally generated dark counts by operating NFADs at lower temperatures compared to those typically achieved with thermoelectric coolers.

Yan et al. \cite{yan2012} have demonstrated the use of NFADs in the free-running regime, operating at a temperature of -80${^o}$C, which achieved a DCR of about 100~cps at 10\% efficiency. However, since there was no active hold-off circuit used, the afterpulse probability was high, preventing operation at higher detection efficiencies. Previously \cite{lunghi2012a}, we developed an NFAD detector with an active hold-off circuit and by operating the devices at -50${^o}$C, a DCR of around 600~cps at 10\% detection efficiency was achieved, whilst the total afterpulse probability could be reduced below 1\% by applying a deadtime of around 10~$\mu$s. 

In this letter, we build on our previous work to investigate the performance of NFADs at lower temperatures, down to -110${^o}$C, which enables operation with extremely low DCR. We also integrate these detectors into a high-speed QKD system which allows high quantum channel loss to be tolerated.

The devices under test were the Princeton Lightwave NFADs (model no. E2G2) which have an active area of 25~$\mu$m and a series quench resistor of 500~k$\Omega$. A description of the electronic readout circuitry used in our detector system can be found in Ref. \cite{lunghi2012a}. Cooling is provided by a free-piston stirling cooler (FPSC) (Twinbird SC-UE15R), which is maintenance free and has a specified cooling power of 20~W at -110${^o}$C. Two NFADs were placed inside a hermetically sealed chamber connected to the cold plate of the cooler, whilst the rest of the electronic circuitry was at room temperature. It was not necessary to evacuate the chamber with a vacuum pump.

In order to ensure that the NFAD is in a well defined initial condition during the characterization, we used a recently developed field programmable gate array (FPGA) based test procedure \cite{lunghi2012a}, which is specially adapted for free-running detectors. The FPGA first imposes a cycle where it waits for no detection to occur in a user defined time, which in this work was set between 75 and 150~$\mu$s, to ensure that the afterpulse traps are emptied. Once this condition is fulfilled, a pulsed laser is triggered and the probability of a detection in the corresponding time-bin provides a measure of the detection efficiency $\eta$. Taking into account the Poisson photon distribution in a laser pulse, $\eta$ is given by

\begin{equation}
\eta=\frac{1}{\mu}\ln  {\left( \frac{1-\frac{r_{dc}}{f}}{1-\frac{C_{d}}{C_{lp}}} \right)} ,
\end{equation}

where the terms are; $\mu$, mean photon number per pulse; $r_{dc}$, dark count rate (measured by disabling the laser); $f$, clock frequency of the FPGA (50 MHz); $C_d$, total number of detection counts in the time-bin synchronized with the laser; $C_{lp}$, total number of laser pulses sent. Conditioned on a detection of the laser pulse, the FPGA looks for an afterpulse and subsequently updates a histogram. Unlike the normal double-window method \cite{cova1991} for gated detectors, this procedure allows the higher-order afterpulse contributions to be easily characterized, i.e. afterpulse of afterpulse. After a sufficient acquisition time, the histogram will reproduce the afterpulsing decay curve, from which the total afterpulse probability, $P_{ap}$, is directly calculated by

\begin{equation}
P_{ap}= \sum_i{\left( \frac{C_i}{C_d}-r_{dc}\tau \right)},
\end{equation}

where $C_i$ is the number of counts occurring in time bin $i$ and $\tau$ is the time bin duration. The afterpulse histograms were recorded for up to 150~$\mu$s after the photon detection. 

\begin{figure}
\includegraphics[width=8.5cm]{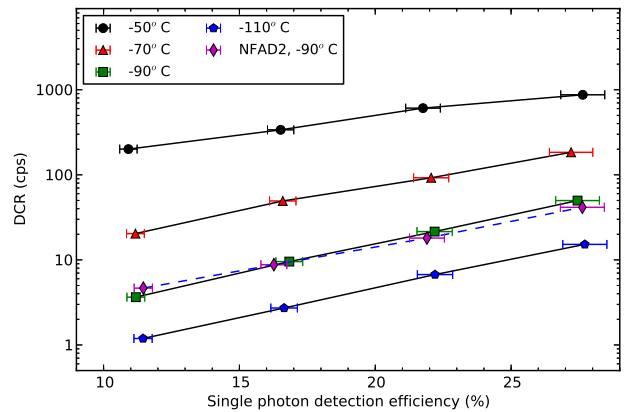}%
\caption{\label{fig:DCR} Dark count rate versus detection efficiency at 1550~nm, for different temperatures between -50$^o$C and -110$^o$C. For comparison, characterization of a second diode (NFAD2) is plotted (dashed line), showing similar performance.}%
 \end{figure}
 
 \begin{figure}
 \includegraphics[width=8.5cm]{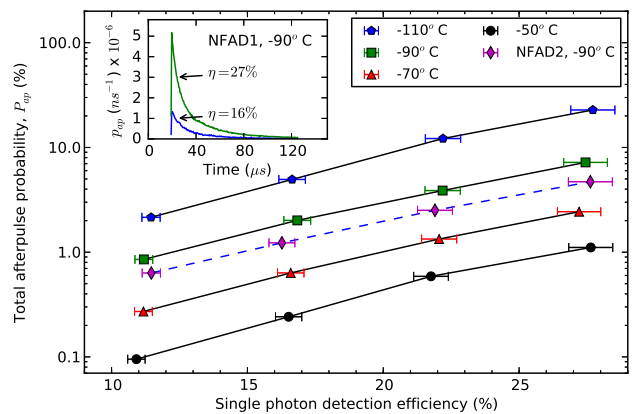}%
 \caption{\label{fig:After} Total afterpulse probability as a function of detection efficiency at different temperatures for 20$\mu$s deadtime. Characterization of a second diode (NFAD2) at -90$^o$C is plotted with the dashed line. Inset shows typical afterpulse histograms (here at -90$^o$C for 16\% and 27\% efficiencies) plotted as an afterpulse probability density ($ns^{-1}$) versus time after a detection. }%
 \end{figure}


Figure \ref{fig:DCR} shows the measured DCR as a function of single photon detection efficiency at 1550~nm for different temperatures between -50${^o}$C and -110${^o}$C, demonstrating detection efficiencies up to 27\%. For this data, the detector hold-off time (deadtime after a detection) was set to 20~$\mu$s. The main source of errors during the efficiency characterization of SPADs is the uncertainty in calculating the number of photons illuminating the detector, which can easily reach 10\%. To ensure high confidence in the detection efficiency measurement we characterized the attenuation in all of the optical components with a calibrated optical power meter, and calculated $\mu=0.91$ with an uncertainty of 2.9\%. Over the temperature range tested, the DCR reduces by over two orders of magnitude, which shows that the dark carrier generation is dominated by a thermally driven Shockley-Read-Hall (SRH) process \cite{Itzler2011}. We expect that the trap-assisted tunnelling limit is near being reached, since the DCR reduction is not constant with each temperature step. At -110${^o}$C, a DCR of 1.19$\pm$0.04~cps was obtained for a detection efficiency of 11.5$\pm$0.3\%, which is nearly two orders of magnitude lower than previously demonstrated for free-running InGaAs detectors. Even at $\eta=$27.7\%, the DCR was only 15.2~cps. 


Figure~\ref{fig:After} shows the total afterpulse probability as a function of detection efficiency, for temperatures between -110$^o$C and -50$^o$C, with a hold-off time of 20~$\mu$s. At the lowest temperature, for an efficiency of 11.5\% the afterpulse probability was 2.2\%, which is acceptable for most applications. Inset of Fig.~\ref{fig:After} shows typical afterpulse histograms for two different detection efficiencies (16\% and 27\%), in this case at a temperature of -90$^o$C, with 20~$\mu$s deadtime. The afterpulse probability increases with decreasing temperature, since the charge carrier trap lifetime increases. Thus, there is a trade off between the DCR and afterpulse probability, for a given efficiency and deadtime.

\begin{figure}
 \includegraphics[width=8.5cm]{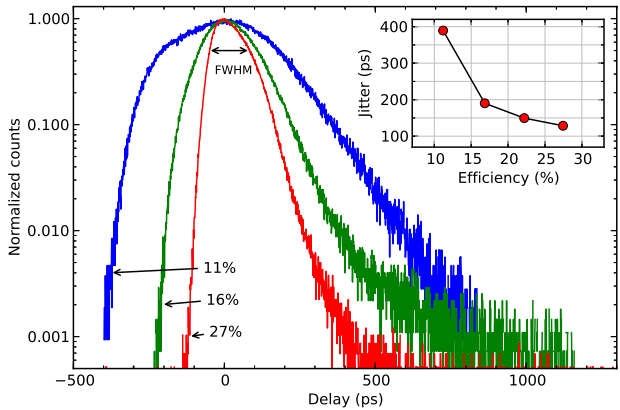}%
 \caption{\label{fig:Jitter} Timing jitter of the detector at -90$^o$C for efficiencies between 11\% and 27\%. Inset shows the FWHM value of the jitter versus detection efficiency.}%
 \end{figure}

The timing jitter of the detector was measured with a time-correlated single photon counting (TCSPC) module, by illuminating the detector with a pulsed laser (PicoQuant) at 1550~nm with a full width at half maximum (FWMH) pulse duration of 33~ps. It was found that below -90$^o$C the jitter did not change significantly with temperature. Figure \ref{fig:Jitter} shows the normalized detector counts as a function of delay from the laser pulse, whilst the inset of Fig.~\ref{fig:Jitter} shows the FWHM jitter ($\Delta t$) as a function of detection efficiency. At $\eta=$27.5\%, $\Delta t$ was 129~ps. The curves are not Gaussian and for many applications we are interested in the jitter at the 1\% level. This is still reasonably small at higher detection efficiencies, e.g. 600~ps at $\eta$=16\%, indicating that these detectors are suitable for high-speed applications such as QKD where detection errors should be minimal.

The best figure of merit to compare single photon detectors for quantum optics applications is given by $H = \eta / ( r_{dc} \Delta t )$, since it takes account of all of these quantities in the most meaningful way \cite{hadfield2009}. At the optimum point of around $\eta = $17\% at -110$^o$C, $H=$ 3.2$\times$10$^8$, which is three orders of magnitude higher than widely used gated InGaAs detectors \cite{eisaman2011}, and becomes comparable with many SNSPD devices.

\begin{figure}
\includegraphics[width=8.5cm]{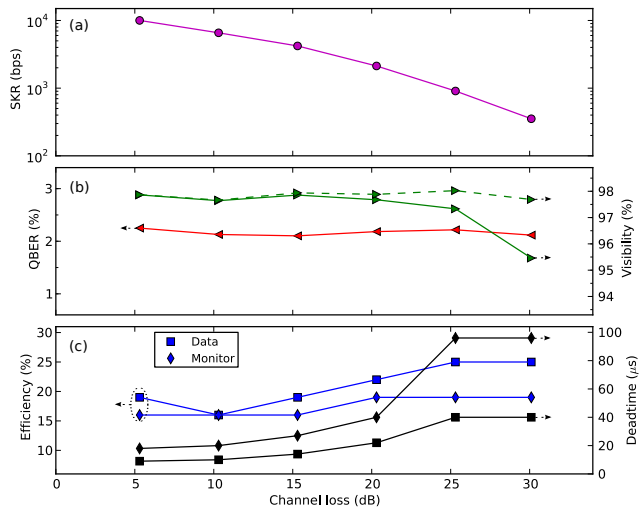}%
\caption{\label{fig:QKD} As a function of channel loss; (a) the secret key rate achieved; (b) raw QBER and visibility (solid lines) and visibility subtracting dark counts in the monitor detector (dashed line); (c) optimized detector efficiency and deadtime, yielding maximum secret key rate.}%
\end{figure}

To demonstrate the stable performance of the detector in a real application, we integrate two NFADs into a high speed QKD system. The system is clocked at 625~MHz and incorporates all of the necessary components for full quantum key distribution, including real-time post-processing for error correction, privacy amplification and classical message authentication \cite{walenta2013b}. The implementation is based on the coherent-one way protocol \cite{stucki2005} which uses two detectors, one for measuring the bit value encoded in the photon arrival time (Data detector) and a second one to measure the visibility of the coherent state interference (Monitor detector), which guarantees security. Due to their extremely low dark count rates, the detectors presented in this letter are especially suited for utilization in QKD scenarios with very long fiber lengths. Moreover, since the photon arrival probability becomes very small, a deadtime on the order of tens of $\mu$s can be tolerated. The link comprised of two 25~km spools of standard single mode fiber (SMF), one for the quantum channel and the second for the bidirectional classical channel. To simulate additional losses in the quantum channel a variable optical attenuator was added. For a given channel loss, there exists an optimum detector temperature for low quantum bit error rate (QBER) and maximum secret key rate, since there is a trade off between dark counts and afterpulsing. For the largest achievable channel loss, we found that the optimum temperature was -90$^o$C. Using security analysis that takes account of finite-key-size effects and a security parameter of $4 \times 10^{-9}$, we obtained secret key rates (SKR) at different channel losses as illustrated in Fig.~\ref{fig:QKD}(a). Correspondingly, Fig.~\ref{fig:QKD}(b) shows the measured QBER and visibility, meanwhile Fig.~\ref{fig:QKD}(c) shows the optimized detection efficiency and deadtime settings for both detectors. With high channel loss, a higher detection efficiency was preferable since a longer deadtime could be tolerated. At lower channel loss the detectors would start to saturate, therefore it was beneficial to reduce the detection efficiency to allow a lower deadtime, which led to higher detection rates. The optimum mean photon number per pulse sent on Alice's side was around 0.06, and the compression ratio for privacy amplification was 15.0\%. With 5~dB of channel loss (25~km of SMF) the secret key rate was 10~kbps. With 30~dB of loss (equivalent to 150~km of SMF) it was still possible to extract 350~bps of secret key, maintaining the same stability, QBER and visibility. The secret key rates presented here are the final output rates from the privacy amplification, after the deduction of secret keys that are used for the classical channel authentication. The hardware key distillation engine \cite{walenta2013b} processes the keys continuously in real-time, without the need for long, individual key sessions. The maximum channel losses presented here are typically not achievable with systems based on InGaAs detectors and would usually require the use of SNSPDs \cite{takesue2007}. We stress that the detectors are currently not the limiting factor in the maximum achievable distance; indeed we believe that it should be possible to increase the channel loss by another 10~dB, provided the system stabilization algorithms, which optimize the QBER and visibility, are adapted to suit lower detection rates. In this scenario the optimum SKR would be achieved at a lower detector temperature, operating with longer deadtime. 

In conclusion we have demonstrated the use of NFADs for free-running detection of single photons at 1550~nm with extremely low DCR, as low as 1~cps at 10\% efficiency, which is an improvement of two orders of magnitude over previously published work \cite{yan2012, lunghi2012a}. This was achieved by utilizing a hold-off circuit to reduce the afterpulse probability and cooling the devices down to -110${^o}$C with a FPSC. We presented the efficiency, dark count, afterpulsing and jitter characterization of the detector as well as a demonstration of two NFADs operating in a high-speed quantum key distribution system. This opens the possibility of using these practical detectors for secure key distribution over distances greater than 150~km. FPSCs are a cheap, compact and convenient cooling solution for temperatures presented in this work. Free-running NFADs provide flexible and robust operation, suitable for many applications in quantum optics, whilst requiring only simple operating electronics.   
 

We acknowledge ID Quantique for providing the electronic detector circuitry for this work; Claudio Barreiro, Raphael Houlmann and Olivier Guinnard for technical support; Mark Itzler for useful discussions; Jacques Morel and Armin Gambon from the Swiss Federal Institute of Metrology for the calibration of the optical power meter. This work was supported by the Swiss National Centre of Competence in Research, Quantum Science and Technology (NCCR-QSIT).

\end{document}